\documentclass[aps,prl,floatfix,twocolumn,superscriptaddress,showpacs]{revtex4}

\pdfoutput=1
\usepackage{graphicx}
\usepackage{amssymb}
\usepackage{amsmath}
\usepackage{color}

\newcommand {\ket}[1] {|#1 \rangle}
\newcommand {\bra}[1] {\langle#1 |}
\newcommand {\braket}[2] {\langle #1 | #2 \rangle}

\newcommand{\comm}[2]{\left[#1,#2\right]}

\newcommand{\vh}{\boldsymbol}

\newcommand{\bea}{\begin{eqnarray}}
\newcommand{\eea}{\end{eqnarray}}
\newcommand{\be}{\begin{equation}}
\newcommand{\ee}{\end{equation}}
\begin{document}

\title{Spontaneous emissions and thermalization of cold bosons in optical lattices}

\author{J.~Schachenmayer} 
  \affiliation{Department of Physics and Astronomy, University of Pittsburgh, Pittsburgh, Pennsylvania 15260, USA}
\author{L.~Pollet} 
\affiliation{Department of Physics, Arnold Sommerfeld Center for Theoretical Physics and Center for NanoScience, University of Munich, 80333 Munich, Germany}
\author{M.~Troyer} 
\affiliation{Theoretische Physik, ETH Zurich, 8093 Zurich, Switzerland}

\author{A.~J.~Daley} \affiliation{Department of Physics and Astronomy, University of Pittsburgh, Pittsburgh, Pennsylvania 15260, USA}

\date{\today}


\begin{abstract}
We study the thermalization of excitations generated by spontaneous
emission events for cold bosons in an optical lattice. Computing the
dynamics described by the many-body master equation, we characterize
equilibration timescales in different parameter regimes. For simple observables,
we find regimes in which the system relaxes rapidly to values
in agreement with a thermal distribution, and others where thermalization
does not occur on typical experimental timescales. Because spontaneous emissions
lead effectively to a local quantum quench, this behavior
is strongly dependent on the low-energy spectrum of the Hamiltonian, and undergoes a qualitative change at the Mott Insulator-superfluid transition point. These results have
important implications for the understanding of thermalization after localized quenches in
isolated quantum gases, as well as the characterization of heating in
experiments.
\end{abstract}

\maketitle 

Spontaneous emission is a fundamental source of heating in optical
dipole potentials \cite{Gordon1980,Dalibard1985}, and one of the key
 heating sources in current experiments with cold atoms in
optical lattices \cite{Gerbier2010,Pichler2010}. 
This heating induces non-equilibrium dynamics in which
thermalization processes are expected to play a key role.
Typically it is assumed that the energy added to the
system will be thermalized, causing an effective increase in temperature. But does that happen?

This question is a special case of a fundamental problem in many-body quantum mechanics: 
to what extent, and under which conditions, will an isolated system undergo
thermalization when perturbed away from equilibrium, in the sense that
at long times the system reaches a steady state where simple
observables take the same values as those of a thermal distribution
\cite{Rigol2008, Cazalilla2011,Rigol2012,Srednicki1994}. Recently, experiments with strongly interacting cold gases 
confined to move in one dimension (1D) \cite{Kinoshita2006} have demonstrated regimes of integrable dynamics - where systems do not thermalize in a traditional sense \cite{Rigol2009}, although they can sometimes relax to a steady-state distribution described by a generalized Gibbs ensemble \cite{Cassidy2011,Rigol2011}. 

In this article we investigate these issues by studying
dynamics induced by spontaneous emissions (incoherent light
scattering) for cold bosons in an optical lattice \cite{Bloch2008}, and identify
contrasting parameter regimes where (i) certain observables relax over short times to 
thermal values, or (ii) the system relaxes on a short timescale to states
that are clearly non-thermal. The dynamics depends greatly on the low-energy spectrum of the Hamiltonian because spontaneous emissions give rise to a local quench, leading to qualitative changes at the superfluid-Mott insulator phase transition. By combining time-dependent density matrix renormalization group
(t-DMRG) methods \cite{Vidal2004,Daley2004,White2004,Verstraete2008}
with quantum trajectory techniques
\cite{Molmer1993,Gardiner2005,Carmichael1993}, we compute the dynamics in the context of 
real experiments. These results 
have important implications for the characterization of heating in
current experiments \cite{McKay2011}. In fact, the lack of thermalization
of certain excitations may be exploited to enhance the realization of fragile
many-body states \cite{Fuchs2011,Jordens2010,Trotzky2010, Esslinger2010}, leading to greater robustness of quantum simulators \cite{Bloch2012,Cirac2012}. Below we first summarize the effects of spontaneous emissions on atoms in an optical lattice, before analyzing thermalization in the lowest Bloch band.

\begin{figure}[tb]
\begin{center}
\includegraphics[width=0.45\textwidth]{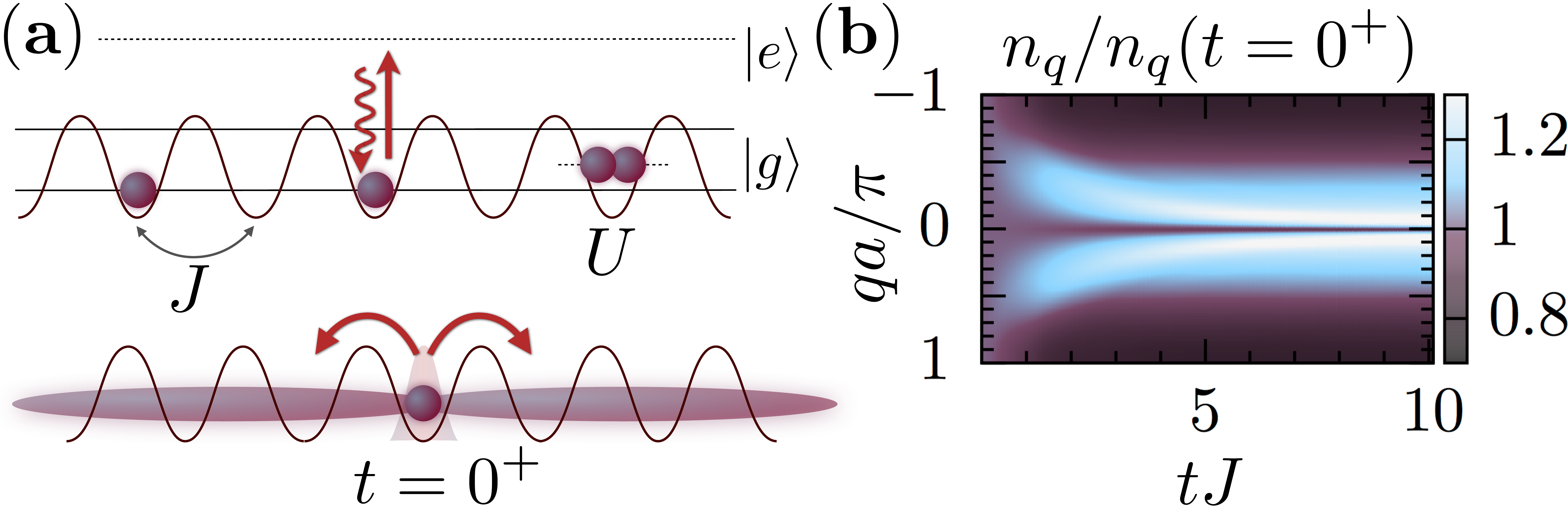}
\caption{(Color online) (a) Absorption and spontaneous emission of a lattice photon
leads effectively to localization of single atoms. Tunnelling and interactions between
atoms then redistribute the energy added to the system. (b) Localization of an atom
in space corresponds initially to a distribution of the atom over the whole Brillouin
  zone (the tails of the quasi-momentum distribution are lifted). Subsequent unitary 
  evolution leads to a broadened quasi-momentum distribution, {\it i.e.}, the $n_{q=0}$ peak and the tails
  decrease, while the small quasi-momentum components increase (t-DMRG, $U=2J$, $N=48$ particles on $M=48$ sites, $d_l=6$,
  $D=512$).\label{fig:1}}
\end{center}
\end{figure}

\emph{Effect of spontaneous emissions --} The scattering of a photon via spontaneous emission effectively provides the environment with information about the position of an atom \cite{Gordon1980,Dalibard1985}. This leads to two key physical effects on bosons beginning in the lowest band of an optical lattice \cite{Gerbier2010,Pichler2010}: It can (i) transfer atoms to higher bands, and/or (ii) localize atoms on the length scale of the photon wavelength $\lambda$.

Transfer of atoms to higher bands is suppressed for the relatively deep optical lattices found in experiments by the square of the Lamb-Dicke parameter, $\eta=2\pi
a_T/\lambda$, where $a_T$ is the trap length for the lowest band Wannier function.  For typical experiments with lattice depths around
$V_0=8E_R$ [with $E_R=4\pi^2 \hbar^2/(2 m \lambda^2)$, where $m$ is
  the mass of the atom], $\eta^2\sim 0.1$, and if the lattice is red-detuned 
  the dominant dissipative processes return atoms to the lowest  band. These rare band transfer processes give rise to a large energy increase of the order of the band-gap energy $\omega_g$ ($\hbar\equiv 1$). This energy is much larger than energy scales in the lowest band, which prevents thermalization of $\omega_g$ on experimental timescales because it would require a collision with many atoms simultaneously to transfer the energy to the lowest band \footnote{Note that collisional processes
  between two or more atoms in the first excited band can return
  particles to the lowest band while exciting atoms to higher
  bands. This doesn't affect the conclusion that the bandgap energy
  cannot be thermalized with the atoms in the lowest
  band.}. This is analogous to the collisional stability of doublon pairs demonstrated in recent experiments \cite{Strohmaier2010}. 
 
 \paragraph{Heating and thermalization in the lowest band-- } For processes where the atom remains in the lowest band, this question is substantially more complicated. A spontaneous emission localizes the atoms on the scale of a single site \cite{Pichler2010}, because the wavelength is comparable to the lattice spacing $\lambda/2 \sim a$. This is in contrast with photon scattering in solid state physics, where $\lambda$ is much larger than the lattice spacing. These processes increase the energy on scales of the width of the lowest band, as atoms are transferred to higher quasi-momentum states. 
 
Thermalization properties then depend on dynamics described by the Bose-Hubbard model,
\bea
H=-J\sum_{\langle \mathbf{i},\mathbf{j}\rangle} b_{\vh{i}}^\dag b_{\vh{j}}+ \frac{U}{2} \sum_\mathbf{i} b_{\vh{i}}^\dag b_{\vh{i}}^\dag b_{\vh{i}}b_{\vh{i}} +\sum_\mathbf{i} \varepsilon_\mathbf{i} b_{\vh{i}}^\dag b_{\vh{i}}. \label{eq:bh}
\eea
Here, $b_{\vh{i}}^\dag$ is a bosonic creation operator for an
atom on site $\vh{i}$, $J$ denotes the tunneling rate between neighboring sites, $U$ the onsite interaction, and $\varepsilon_\mathbf{i}$ the onsite potential. This model is non-integrable outside the limiting cases of $U\rightarrow 0$ and $U/J \rightarrow \infty$, and has been shown to exhibit chaotic spectral properties when $U\sim J$ \cite{0295-5075-68-5-632,PhysRevLett.99.020401}. As a result, it might be expected that the system will thermalize for most values of $U/J$, with the most rapid thermalization around $U\sim J$. For high values of $U/J$, the  system behaves as hard-core bosons, relaxing to a generalized Gibbs ensemble \cite{Rigol2007,Rigol2009}. This is what is typically expected for a global quench of the value of $U/J$ \cite{Biroli2010}. However, it is not clear that this analysis applies to our situation because a spontaneous emission event leads to localization of atoms in a \emph{local quantum quench} with excitations that are very low in energy. Because the lowest part of the energy spectrum can exhibit spectral statistics closer to an integrable model \cite{0295-5075-68-5-632}, this may even result in a lack of thermalization for all values of $U/J$. Below we find that the relaxation timescales and equilibrium values strongly depend on the interactions in the lower band (as it is also observed for local quenches in 2D \cite{Natu2011,Hung2010,Bakr2010}).

In the lowest band, the heating and thermalization together can be effectively described by a master equation \cite{Pichler2010} (see supplementary material), 
\bea
\dot\rho=-i\comm{H}{\rho}-\frac{\gamma}{2}\sum_{\mathbf{i}}
\comm{\hat n_{\vh{i}}}
{\comm{\hat n_{\vh{i}}}{\rho}}, \label{eq:banddissipative}
 \eea
where $H$ is the Bose-Hubbard Hamiltonian \eqref{eq:bh}. The dissipative dynamics involve localization of particles on a single site via scattering of photons at a rate $\gamma$, which depends on the intensity of the lattice lasers and the detuning from resonance.

\paragraph{Thermalization after a single intra-band spontaneous emission
  event --} In order to characterize the thermalization process, we first consider the
situation where the system is in the ground state of model  \eqref{eq:bh} $\ket{\psi_g}$ at time $t=0$, and  undergoes a
spontaneous emission (on site $i$). In the sense of continuous
measurement theory \cite{Gardiner2005} applied to \eqref{eq:banddissipative}, the resulting state prepared is 
\begin{equation}
\ket{\psi_\mathbf{i}(t=0^+ )}=\frac{\hat n_\mathbf{i}\ket{\psi_g}}{||\hat n_\mathbf{i}\ket{\psi_g}||}.
\label{eq:single_spont_em}
\end{equation}
We consider a weighted ensemble average over the sites $\mathbf{i}$ with probabilities of spontaneous emission $p_\mathbf{i}\propto \langle \psi_g|n_\mathbf{i}^2 |\psi_g\rangle$, and treat a 1D system, where we can use t-DMRG methods to propagate the state exactly in time. Note that all t-DMRG results are converged in the matrix product state bond dimension $D$ and the truncation of the local dimension, $d_l$.

\begin{figure}[tb]
\begin{center}
\includegraphics[width=0.48\textwidth]{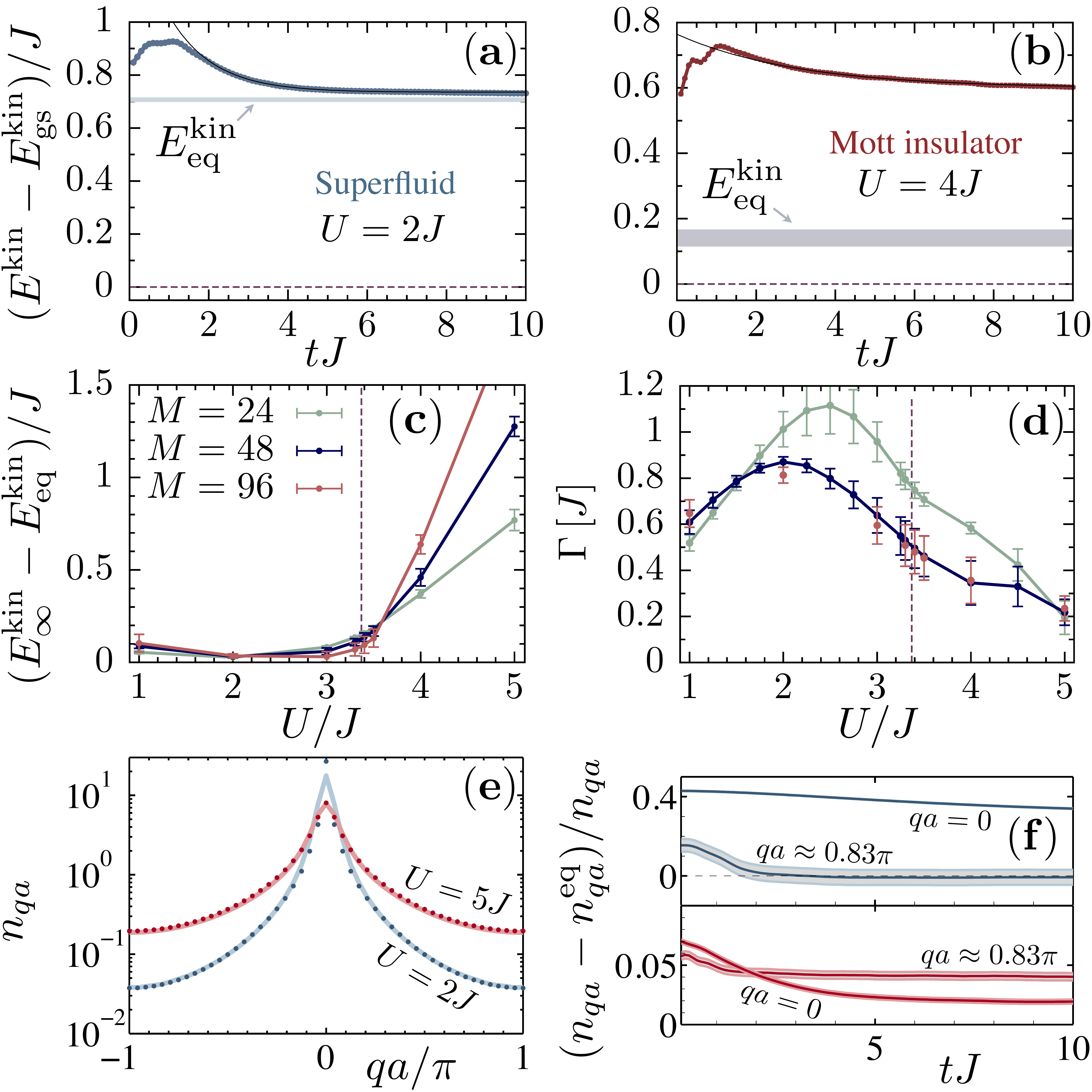}
\caption{(Color online) Time evolution after a single spontaneous
emission. (a - b) For
a superfluid initial state ($U=2J$), the kinetic energy relaxes
to the equilibrium value obtained from a Monte-Carlo calculation
$E_{\rm eq}^{\rm kin}$. For MI states ($U=4J$), the energy
relaxes, but not to $E_{\rm eq}^{\rm kin}$. The zero value of
kinetic energy for this plot is the ground state kinetic energy
$E_{\rm gs}^{\rm kin}$. (c) The difference of the infinite time
value of the kinetic energy (obtained from an extrapolation of an
exponential fit) to the equilibrium energy. For MI
states with $U/J \gtrsim 3.37$, the difference increases rapidly
for $M=24,48,96$ sites (d) The decay rate extracted
from the exponential fit as a function of $U$. (e) Comparison of
the time-evolved quasi-momentum distribution at $t=10/J$ (dots)
to the equilibrium distribution from a QMC calculation. (f/g)
Differences between the two distributions as a function of time
for the $qa=0$ peak and for a large quasi-momentum of $qa=(40/48) \pi$. In
the superfluid [$U=2J$, (f)], the components for large momenta relax rapidly to thermal values, for $qa\sim 0$, the relaxation timescale is much longer. In the MI [$U=5J$, (g)], the same is true, but for large momenta there is a discrepancy to the thermal value.
(t-DMRG, $d_l=6$, $D=256,512$; error bars represent
fitting errors and statistical errors from QMC). \label{fig:2}}
\end{center}
\end{figure}

Fig.~1b shows the typical dynamics after a spontaneous emission spreads a
particle over the whole Brillouin zone and increases the kinetic
energy $E^{\rm kin}$. The interactions between
particles transfer some of this increased kinetic energy to
interaction energy, as shown explicitly in Fig.~2a for an
initial superfluid (SF) state with $U=2 J$. At $t=0^+$, $E^{\rm kin}$ is increased by an amount of the order of $J$ over the ground state value, and it then relaxes to lower value over a timescale $\sim 5/J$ in unitary time evolution. We obtain an equilibrium value $E^{\rm kin}_{\rm eq}$ from path integral Monte-Carlo (QMC) calculations with worm-type updates~\cite{Prokofev1998} (here in the implementation of Ref.~\cite{Pollet2007} -- see Ref.~\cite{Pollet2012} for a recent review of the method with applications to cold gases) at finite temperature $T$,
fitting $T$ to match the value of energy $\langle E \rangle$ for $t\geq 0^+$. It is remarkable that this value corresponds to the equilibrium value reached
dynamically within statistical errors, indicating thermalization of this quantity. In contrast, for an initial Mott Insulator (MI, $U=4J$) state,
$E^{\rm kin}$ relaxes on a slightly longer timescale to an equilibrium value that clearly does not correspond to a thermal distribution at the
appropriate value of $\langle E \rangle$. In fact, in this parameter regime, 
thermally induced coherence in the MI leads to a $E^{\rm kin}_{\rm eq}$ being close or even  below the value of
the ground state kinetic energy \cite{Toth2011}. 

In Fig.~2c we compare the extrapolated equilibrium kinetic energy, $E_{\infty}^{\rm kin}$ (obtained from an exponential fit) to $E^{\rm kin}_{\rm eq}$ for various system sizes and interaction strengths. The lack of thermalization for values of $U/J$ immediately above the SF-MI transition point (when the gap is about $\Delta = J/8$)
is  striking. Although from our calculations we cannot rule out a second relaxation process to a thermal distribution for much larger systems or on much longer timescales, it is clear that a qualitative change in behavior occurs here, leading to a lack of thermalization on typical experimental timescales. Before performing these calculations, we might have expected a crossover behavior, similar to that seen in the relaxation rates, as shown in Fig.~2d from exponential
fits to the long-time behavior of $E^{\rm kin}$, where the fastest relaxation occurs for $U/J\sim 1$. 

\begin{figure}[tb]
\begin{center}
\includegraphics[width=0.48\textwidth]{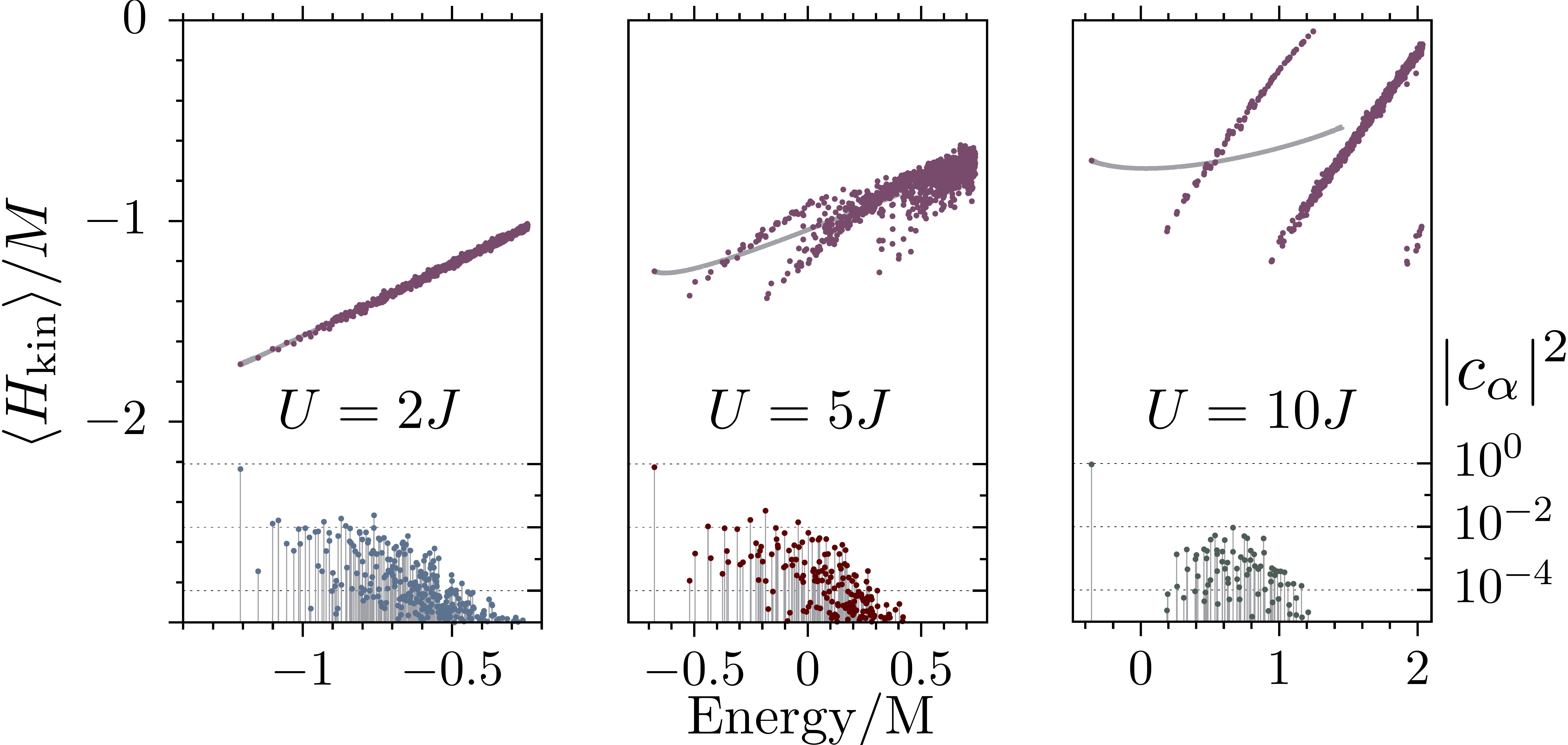}
\caption{(Color online) Expectation values of the kinetic energy of the lowest $1000$ eigenstates as a function of the energy in a system with $M=10$ and $N=10$  (exact diagonalization).  The grey line in the upper plot shows the equilibrium kinetic energy $E^{\rm kin}_{\rm eq}$ for increasing temperatures as a function of the mean energy of the underlying Boltzmann distributions. In the SF, the eigenvalue expectations are distributed around $E^{\rm kin}_{\rm eq}$, but are far from these values in the MI. The lower parts show the occupation probabilities for eigenstates after a single spontaneous emission. \label{fig:3} }
\end{center}
\end{figure}

Note that as with thermalization in any closed quantum system, the behavior depends on the observable considered, and sufficiently complicated or non-local observables will never thermalize \cite{Rigol2008}. In Figs.~2e,f, we show the quasi-momentum distribution $n_q$ in our system with open boundary conditions for different points in time. For all $q$ except very near $q=0$, $n_q$ relaxes to a thermal distribution on timescales $tJ\sim 5$ in the SF for $U\gtrsim 1$. However, long wavelength modes around $n_{q=0}$ require much longer relaxation timescales, and are still far from their steady state values on the timescales computed here (though they are evolving towards the expected thermal value). In the MI, the distribution behaves qualitatively differently, in that all values of $q$ show small discrepancies from the equivalent thermal values, consistent with what we observed for the kinetic energy. While these discrepancies are small for a single spontaneous emission event, they can be much larger when multiple photons are scattered in the experimental protocol discussed below. 

\emph{Explanation based on the low-energy spectrum --} The key to understanding the qualitative change in behavior at the transition point lies in the fact that the spontaneous emissions give rise to a local quantum quench, which only significantly populates low-energy eigenstates. Most of the amplitude of the resulting wavefunction is in the ground state ( Fig.~\ref{fig:3}), where we plot occupation probabilities $|c_\alpha|^2$ and expectation values of the kinetic energy $\langle E_\alpha |  \hat E^{\rm kin} |E_\alpha\rangle$ in the lowest $1000$ energy eigenstates $| E_\alpha \rangle$. We find that $E^{\rm kin}$ grows essentially linearly as a function of $E_\alpha$, even for $U/J\sim 3$ near the phase transition, and that these values coincide with $E^{\rm kin}_{\rm eq}$ from Boltzmann distributions with corresponding  mean energies $E_\alpha$. Therefore, a state with $|c_\alpha|^2 $ leading to an energy expectation $\langle E \rangle$ will approximately have the same kinetic energy as $E^{\rm kin}_{\rm eq}$ with mean energy $\langle E \rangle$. Thus, also the long time average $\langle E^{\rm kin}\rangle \rightarrow \sum_\alpha |c_\alpha |^2 \langle E_\alpha |  \hat E^{\rm kin} |E_\alpha\rangle $ \cite{Rigol2008} will correspond to $E^{\rm kin}_{\rm eq}$ for the corresponding $\langle E \rangle$. As soon as we enter the MI phase, between $U/J\approx 3$ and $U/J \approx 3.8$, there is a qualitative change in the distribution of $\langle E_\alpha |  \hat E^{\rm kin} |E_\alpha\rangle$, as depicted in Fig.~\ref{fig:3}, after which we cannot expect to obtain thermal values. In the deep MI, $\langle E_\alpha |  \hat E^{\rm kin} |E_\alpha\rangle$ are far from $E^{\rm kin}_{\rm eq}$, and correspond to excitations of doublon-holon pairs. In this limit, the system will relax over time to a generalized Gibbs ensemble.


\begin{figure}[tb]
\begin{center}
\includegraphics[width=0.48\textwidth]{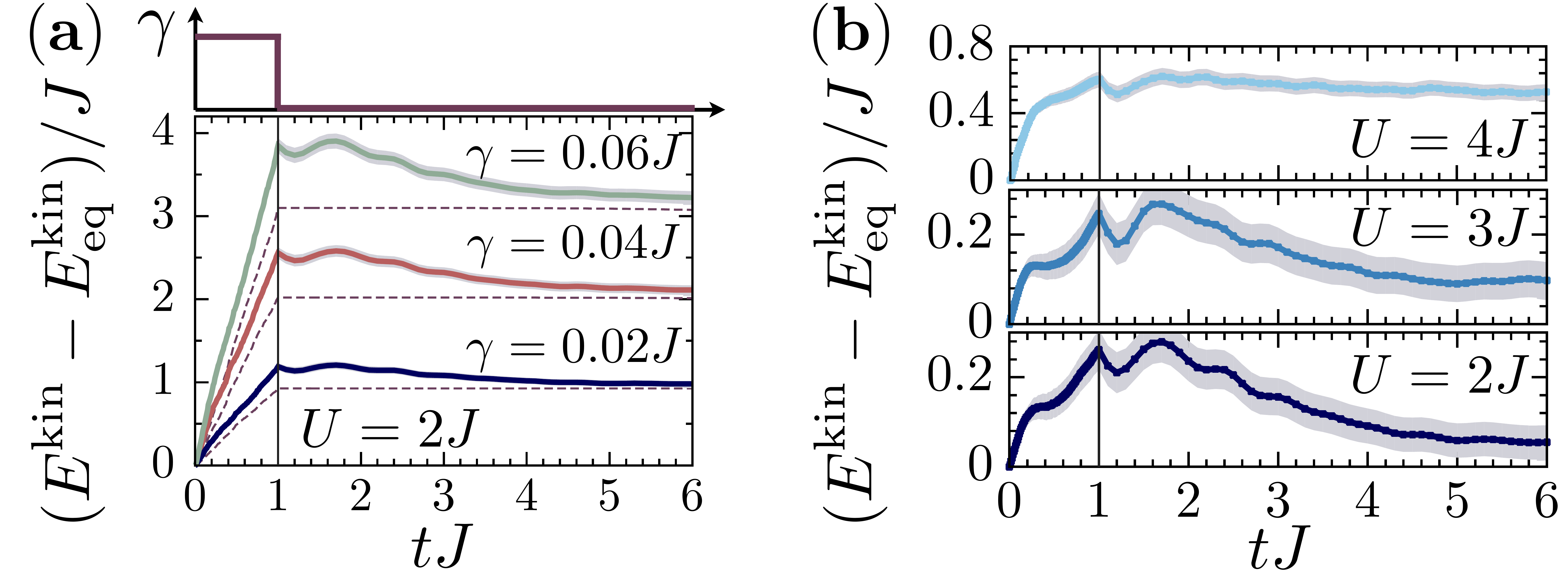}
\caption{(Color online) Quantum trajectory simulations for heating with spontaneous
emission rate $\gamma$, which is switched on for a time $t=1/J$,
as illustrated above panel (a). $M=N=48,$ and the
standard error of the mean is given as shaded area. (a) The
increase in kinetic energy during the heating and the subsequent
relaxation for $\gamma=0.02,0.04,0.06$. For superfluid initial
states, the kinetic energy relaxes to the equilibrium value (QMC
calculations, dashed lines). For a Mott insulating initial state,
on the same time-scale, the energy does not thermalize. This can
be seen in (b) where we plot the difference between the actual
kinetic energy and the equilibrium energy (t-DMRG results,
$D=256$, $d_l=6$, $500$ trajectories). \label{fig:4}}
\end{center}
\end{figure}

\paragraph{Proposed experimental measurement -- } We now consider a specific experimental setup in
which these effects could be observed. It is important to consider multiple spontaneous emission events, both because of the difficulty of restricting to a single event, and in order to make the change in the momentum distribution sufficiently large to measure. As depicted in Fig.~\ref{fig:4}a, we
consider a situation in which the background scattering rate is low, and then a moderate
scattering rate is induced for a short time $t=1/J$ ({\it e.g.}, via a weak
beam with near-resonant light). We then switch this off, and observe
how the system thermalizes over a timescale of $t\sim 5/J$. We compute
the dynamics of this process by combining t-DMRG methods with quantum
trajectory techniques \cite{Daley2009}, which after a stochastic average allow us to
determine the many-body dynamics from the master equation (see supplementary material).

In Fig.~4a, we plot
$E^{\rm kin}$ and $E^{\rm kin}_{\rm eq}$ as a function of time.  As expected from our single-event calculations, the $E^{\rm kin}$ increases much faster than would
be expected from a thermal distribution with the same increase in
total energy (dashed line), and this is more pronounced for larger $\gamma$. Note that in the experiment of
Ref.~\cite{Trotzky2010}, $\gamma\approx 0.02 J$. In Fig.~4b we plot $E^{\rm kin}-E^{\rm kin}_{\rm eq}$ for
different values of $U/J$. We see clearly that as in the case of a
single spontaneous emission,  the
kinetic energy will relax towards the expected equilibrium
values in the superfluid regime. Strikingly, this is not the case in the Mott Insulator,
where the system remains well away from the equilibrium value on the
timescales calculated. Note that while here the energy increase is small, as we use parameters
where few spontaneous emission events occur to allow quantitative numerical treatments, 
experiments could work with faster scattering rates or longer
excitation timescales. Our predictions are observable via momentum
distribution measurements that study relaxation in different parameter regimes. This would be enhanced by a quantitative comparison between experimental measurements and QMC calculations (similar to Ref.~\cite{Trotzky2010}).

\paragraph{Conclusions} --  We showed that for bosons in an optical lattice, a change in the thermalization behavior after spontaneous emissions occurs at the SF-MI transition point. Simple quantities including the kinetic energy and quasi-momentum distribution settle rapidly to a steady state. However, while in some cases these values correspond to a thermal distribution, in others the values are demonstrably non-thermal. These findings, presented here for a uniform system, remain valid in the presence of a harmonic trap, as is shown by results presented in the supplementary material. The lack of complete thermalization implies that the specific effects on specific many-body states must be considered. The generalization of these results to higher dimensions remains an open question, however, because this is a low-energy quench, we expect also that the thermalization properties will be strongly dependent on the detailed low-energy spectrum.

In some regimes, this may lead to greater robustness of states produced in optical lattices, especially where the energy added in a spontaneous emission event would correspond to temperatures above those required for realization of fragile types of order \cite{Fuchs2011,Jordens2010,Trotzky2010, Esslinger2010}. Because the dynamics must instead be treated as a non-equilibrium situation on a case-by-case basis, much of the interesting order can survive on significant timescales.

We thank I.~Bloch, D.~Boyanovsky, W.~Ketterle, S.~Langer, H.~Pichler,
U.~Schneider, D.~Weiss, and P.~Zoller for helpful and motivating
discussions. This work was supported in part by AFOSR grant
FA9550-13-1-0093,  by a grant from the US Army Research Office with
funding from the DARPA OLE program. We acknowledge hospitality of the Aspen Center for Physics, supported by NSF grant PHY-1066293.
  Computational resources were
provided by the Center for Simulation and Modeling at the University
of Pittsburgh.

\newpage
\begin{widetext}
\section*{SUPPLEMENTARY MATERIAL \\Spontaneous emissions and thermalization of cold bosons in optical lattices}

\section{Origin of the description of spontaneous emissions}
In Ref.~\cite{Pichler2010}, a many-body master equation was derived to
describe the effects of spontaneous emission processes for bosonic atoms in an optical lattice. For far-detuned optical
fields, this can be obtained by adiabatically eliminating the excited atomic levels, obtaining
 an effective equation for ground-state atoms. When the lattice
spacing $a$ is comparable to or greater than the optical wavelength of
scattered photons, $a\gtrsim \lambda$, the dynamics of the
many-body density operator $\rho$ is ($\hbar\equiv 1$), $\dot\rho=-i\comm{H}{\rho}+\mathcal{L}_1\rho$, where the dissipative term describing scattering of laser
photons, denoted $\mathcal{L}_1\rho$ is
\bea
\mathcal{L}_1\rho&=&-\frac{1}{2}\sum_{\mathbf{klmni}}
\gamma_{\vh{k}\vh{l}\vh{m}\vh{n}}\comm{b_{\vh{i}}^{(\vh{k})\,\dag}b_{\vh{i}}^{(\vh{l})}}
{\comm{b_{\vh{i}}^{(\vh{m})\,\dag}b_{\vh{i}}^{(\vh{n})}}{\rho}}, \label{eq:multibanddissipative}
 \eea
and $H$ is a multi-band Bose-Hubbard Hamiltonian \cite{Pichler2010},
\begin{align}
H&=-\sum_{\vh{n},\langle\vh{i},\vh{j}\rangle}J^{(\vh{n})}b_{\vh{i}}^{(\vh{n})\,\dag}b_{\vh{j}}^{(\vh{n})} + \sum_{\vh{n},\vh{i}}\varepsilon_{\vh{i}}^{(\vh{n})}b_{\vh{i}}^{(\vh{n})\,\dag}b_{\vh{i}}^{(\vh{n})}\\
&+\sum_{\vh{i},\vh{k},\vh{l}\vh{m},\vh{n}}\frac{1}{2}U^{(\vh{k,l,m,n})}b_{\vh{i}}^{(\vh{k})\,\dag}b_{\vh{i}}^{(\vh{l})\,\dag}b_{\vh{i}}^{(\vh{m})}b_{\vh{i}}^{(\vh{n})}.
\label{eq:multibandham}
\end{align}
Here, the 3D band indices are denoted by $\mathbf{k,l,m,n}$, and
$b_{\vh{i}}^{(\vh{n})\,\dag}$ is a bosonic creation operator for an
atom on site $\vh{i}$ in band $\vh{n}$. The dissipative dynamics
involves scattering of photons and (in some cases) transitions between
Bloch bands with the corresponding rates denoted
$\gamma_{\vh{k}\vh{l}\vh{m}\vh{n}}$, while the tunnelling in band
$\vh{m}$ between neighboring sites $\langle i,j \rangle$ is denoted
$J^{(\vh{n})}$, on-site interactions are denoted $U^{(\vh{k,l,m,n})}$
and an onsite potential $\varepsilon_{\vh{i}}^{(\vh{n})}$. Note that
this includes the band energy $\omega^{(\vh{n})}$ as well as
(potentially) an external trapping potential. The parameters can be calculated from the microscopic model by expanding in a 
basis of Wannier functions \cite{Pichler2010}, though care must be taken to use properly
regularized potentials in evaluating the interaction matrix elements $U^{(\vh{k,l,m,n})}$ \cite{Buchler2010,Mark2011,Mark2012}.

All of the coefficients $\gamma_{\vh{k}\vh{l}\vh{m}\vh{n}}$ depend on the intensity of the laser light via the effective Rabi frequency $\Omega$ and also on the detuning from atomic resonance $\Delta$. In the approximation of a two-level atom, $\gamma \propto \Omega^2/\Delta$. As discussed in the main text, in deep optical lattices, transition rates $\gamma_{\vh{k}\vh{l}\vh{m}\vh{n}}$ for \textit{inter-band} processes coupling neighboring Bloch bands are suppressed by the square of the Lamb-Dicke parameter, $\eta=2\pi
a_T/\lambda$, where $a_T$ is the trap length for the lowest band Wannier function.  For typical experiments with lattice depths around
$V_0=8E_R$ [with $E_R=4\pi^2 \hbar^2/(2 m \lambda^2)$, where $m$ is
  the mass of the atom], the suppression is $\eta^2\sim 0.1$.
   In the usual case of a red-detuned optical lattice
  the dominant dissipative processes are thus \textit{intra-band} processes, which return the atoms to their initial Bloch band. Processes
  accessing higher Bloch bands are suppressed by a factor of the order $\eta^4$ or greater, and we can
write an effective two-band master equation describing the dynamics of
the density operator for atoms in the lattice. In the main text, we set $\gamma_{\vh{0}\vh{0}\vh{0}\vh{0}}\equiv \gamma$, so that in
the Lamb-Dicke limit $\eta\ll 1$, $\gamma_{\vh{1}\vh{0}\vh{1}\vh{0}}=\gamma_{\vh{0}\vh{1}\vh{0}\vh{1}}=\eta^2\gamma$. We also use the symbols $U\equiv U^{(\vh{0,0,0,0})}$ and
$J=J{(\vh{0})}$, in order to obtain eq.~(1) of the main text.

\section{Effects of a harmonic trap}

In a realistic
experimental setup, the particles will always be confined by a
harmonic trap. This can lead to situations, in which parts of the
system have superfluid components despite the fact that $U/J >
3.37$. In Fig.~\ref{fig:a1}b we show results for the evolution after a single
spontaneous emission in the presence of a harmonic confinement
$\varepsilon_i=\omega_T \sum_i (i-i_0)^2$ (center site $i_0=(M+1)/2$) with
$\omega_T=0.012J$ for a system with $M=48$ sites. We again average over jumps on all possible sites weighted with the probabilities $p_i \propto \langle \psi_g | n_i^2 | \psi_g \rangle $. In the upper panel
we find thermalization for the kinetic energy in the superfluid state with $U=2J$. As in the
case of box boundary conditions, the kinetic energy relaxes to a value
corresponding to a thermal ditribution on the experimentally relevant
time-scales. 
For larger interactions ($U=5J$), we see from the density
profile (insets) that the system is not  in a  MI state with unit filling
(these appear only for values around $U/J\sim 10$). Nevertheless, in
this case we find that the system does not relax to a value of a thermal distribution. Note that the
Monte-carlo thermal kinetic energy in this case is in fact above the
value after the jump. 
In the case of a Mott insulator in a trap ($U=10J$) with unit filling
(seen by the density profile in the inset), we find that the system
does not approach a steady state kinetic energy but shows
oscillations, which can be explained by boundary effects of deflected doublon-holon pairs.

\begin{figure}[tb]
\begin{center}
\includegraphics[width=0.68\textwidth]{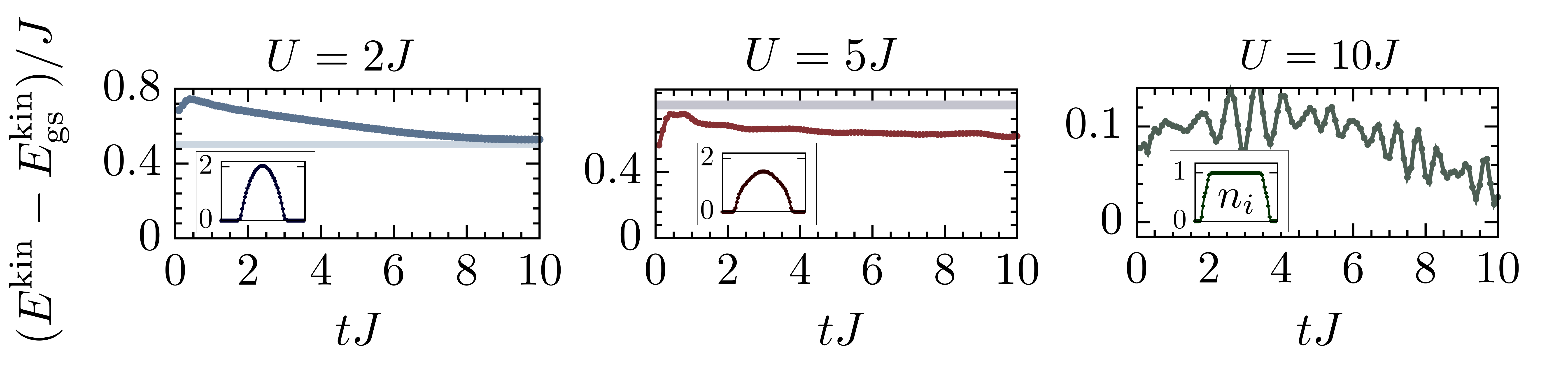}
\caption{(Color online) Relaxation in the
  lowest band in the presence of a harmonic trap
  ($\omega_T^2=0.012J$, $N=48$, $M=64$). The insets show the initial
  density profiles of the ground-states in the trap. (t-DMRG results, $D=512$, $d_l=6,8$)  \label{fig:a1} }
\end{center}
\end{figure}

\section{Details of numerical calculations}

For single spontaneous emission events, we find that t-DMRG methods are very efficient. Specifically because of the form of the local quench, we find that we can obtain converged numerical results in which truncation errors are unimportant up to $tJ\sim 10$. Beyond that, it is difficult to get exceptionally accurate results using t-DMRG, because truncation errors grow as a function of time.

To simulate the time evolution of the finite-light time scattering presented in the manuscript we make use of a quantum-trajectory technique \cite{carmichael_open_1991,molmer_monte_1993,dum_monte_1992} in combination with time-dependent Density Matrix Renormalization Group (t-DMRG) methods Refs. \cite{Vidal2004,Daley2004,White2004}. Therefore we write the master equation as ($\hbar\equiv 1$)
\begin{align}
	\label{eq:qt_meq}
        \frac{d}{dt} \hat \rho = 
        -i \left( \hat H_{\rm eff} \hat \rho
        - \hat \rho \hat H_{\rm eff} \right)
        + \sum_\mu \hat n_\mu \hat \rho \hat n_\mu
\end{align}
with the effective non-hermitian Hamiltonian
\begin{align}
    \hat H_{\rm eff}= \hat H - \frac{i}{2} \sum_\mu  
     \hat n_\mu\hat n_\mu.
\end{align}
Instead of integrating Eq.~\eqref{eq:qt_meq}, the idea of quantum trajectories is to approximate the time-dependent expectation value of any observable $\hat O$, for an initial pure density matrix $\rho(t=0)=\ket{\psi_{\rm in}} \bra{\psi_{\rm in}}$ as
\begin{align}
\label{eq:traj}
 \langle \hat O \rangle = {\rm tr} ( \hat O \rho(t)) 
\approx \frac{1}{\mathcal{M}} \sum_n^{\mathcal{M}} \bra{\psi^{[n]}(t)} \hat O  \ket{\psi^{[n]}(t)}
.
\end{align}
Here $\mathcal{M}$ is a large number of "trajectories". It can be shown \cite{carmichael_open_1991,molmer_monte_1993,dum_monte_1992}  that for $\mathcal{M}\to \infty$, Eq.~\eqref{eq:traj} becomes exact if the state for the $n$-th trajectory at time t,  $\ket{\psi^{[n]}(t)}$ is calculated via a probabilistic evolution of the following form: i) Draw  a uniform random number $r \in [0,1]$ and evolve the state via   $\ket{\phi^{[n]}}={\rm exp} (-itH_{\rm eff}  t^*) \ket{\psi_{\rm in}}$ until the norm, which decreases due to the coherent evolution with the non-hermitian Hamiltonian, drops below $q_l=|\braket{\phi^{[n]}}{\phi^{[n]}}|^2 < r$ at time $t^*$; ii) Apply a quantum jump at a site $\mu$ according to a probability distribution  $p_\mu \propto \bra{\psi(t^*)}  \hat n_\mu \hat n_\mu \ket{\psi(t^*)} $; iii) Choose the normalized state $ \ket{\psi^{[n]}_{\rm in}}= \hat n_\mu \ket{\psi(t^*)} / \| \hat n_\mu \ket{\psi(t^*)}\|$ as new input state for (i) with a new random number $r$. iii) When time $t$ is reached, normalize it and use it to calculate the expectation value in Eq.~\eqref{eq:traj}.

Collecting many trajectories makes it possible to obtain a good estimate for  $\langle \hat O \rangle$. The coherent evolution with non-hermitian Hamiltonians can be implemented completely analogously to standard t-DMRG techniques. The application of a local jump operator to a matrix product state is straightforward. To obtain correct results, besides the time-step the convergence in the bond-dimension for each trajectory has to be checked.

\vspace{5cm}

   \def\eprint#1{arXiv:#1}
   
   \end{widetext}

\bibliographystyle{apsrev}
\bibliography{therm}

\begin{thebibliography}{46}
\expandafter\ifx\csname natexlab\endcsname\relax\def\natexlab#1{#1}\fi
\expandafter\ifx\csname bibnamefont\endcsname\relax
  \def\bibnamefont#1{#1}\fi
\expandafter\ifx\csname bibfnamefont\endcsname\relax
  \def\bibfnamefont#1{#1}\fi
\expandafter\ifx\csname citenamefont\endcsname\relax
  \def\citenamefont#1{#1}\fi
\expandafter\ifx\csname url\endcsname\relax
  \def\url#1{\texttt{#1}}\fi
\expandafter\ifx\csname urlprefix\endcsname\relax\def\urlprefix{URL }\fi
\providecommand{\bibinfo}[2]{#2}
\providecommand{\eprint}[2][]{\url{#2}}

\bibitem[{\citenamefont{Gordon and Ashkin}(1980)}]{Gordon1980}
\bibinfo{author}{\bibfnamefont{J.~P.} \bibnamefont{Gordon}} \bibnamefont{and}
  \bibinfo{author}{\bibfnamefont{A.}~\bibnamefont{Ashkin}},
  \bibinfo{journal}{Phys. Rev. A} \textbf{\bibinfo{volume}{21}},
  \bibinfo{pages}{1606} (\bibinfo{year}{1980}).

\bibitem[{\citenamefont{Dalibard and Cohen-Tannoudji}(1985)}]{Dalibard1985}
\bibinfo{author}{\bibfnamefont{J.}~\bibnamefont{Dalibard}} \bibnamefont{and}
  \bibinfo{author}{\bibfnamefont{C.}~\bibnamefont{Cohen-Tannoudji}},
  \bibinfo{journal}{Journal of Physics B: Atomic and Molecular Physics}
  \textbf{\bibinfo{volume}{18}}, \bibinfo{pages}{1661} (\bibinfo{year}{1985}).

\bibitem[{\citenamefont{Gerbier and Castin}(2010)}]{Gerbier2010}
\bibinfo{author}{\bibfnamefont{F.}~\bibnamefont{Gerbier}} \bibnamefont{and}
  \bibinfo{author}{\bibfnamefont{Y.}~\bibnamefont{Castin}},
  \bibinfo{journal}{Phys. Rev. A} \textbf{\bibinfo{volume}{82}},
  \bibinfo{pages}{013615} (\bibinfo{year}{2010}).

\bibitem[{\citenamefont{Pichler et~al.}(2010)\citenamefont{Pichler, Daley, and
  Zoller}}]{Pichler2010}
\bibinfo{author}{\bibfnamefont{H.}~\bibnamefont{Pichler}},
  \bibinfo{author}{\bibfnamefont{A.~J.} \bibnamefont{Daley}}, \bibnamefont{and}
  \bibinfo{author}{\bibfnamefont{P.}~\bibnamefont{Zoller}},
  \bibinfo{journal}{Phys. Rev. A} \textbf{\bibinfo{volume}{82}},
  \bibinfo{pages}{063605} (\bibinfo{year}{2010}).

\bibitem[{\citenamefont{Rigol et~al.}(2008)\citenamefont{Rigol, Dunjko, and
  Olshanii}}]{Rigol2008}
\bibinfo{author}{\bibfnamefont{M.}~\bibnamefont{Rigol}},
  \bibinfo{author}{\bibfnamefont{V.}~\bibnamefont{Dunjko}}, \bibnamefont{and}
  \bibinfo{author}{\bibfnamefont{M.}~\bibnamefont{Olshanii}},
  \bibinfo{journal}{Nature} \textbf{\bibinfo{volume}{452}},
  \bibinfo{pages}{854} (\bibinfo{year}{2008}).

\bibitem[{\citenamefont{Cazalilla et~al.}(2011)\citenamefont{Cazalilla, Citro,
  Giamarchi, Orignac, and Rigol}}]{Cazalilla2011}
\bibinfo{author}{\bibfnamefont{M.~A.} \bibnamefont{Cazalilla}},
  \bibinfo{author}{\bibfnamefont{R.}~\bibnamefont{Citro}},
  \bibinfo{author}{\bibfnamefont{T.}~\bibnamefont{Giamarchi}},
  \bibinfo{author}{\bibfnamefont{E.}~\bibnamefont{Orignac}}, \bibnamefont{and}
  \bibinfo{author}{\bibfnamefont{M.}~\bibnamefont{Rigol}},
  \bibinfo{journal}{Rev. Mod. Phys.} \textbf{\bibinfo{volume}{83}},
  \bibinfo{pages}{1405} (\bibinfo{year}{2011}).

\bibitem[{\citenamefont{Rigol and Srednicki}(2012)}]{Rigol2012}
\bibinfo{author}{\bibfnamefont{M.}~\bibnamefont{Rigol}} \bibnamefont{and}
  \bibinfo{author}{\bibfnamefont{M.}~\bibnamefont{Srednicki}},
  \bibinfo{journal}{Phys. Rev. Lett.} \textbf{\bibinfo{volume}{108}},
  \bibinfo{pages}{110601} (\bibinfo{year}{2012}).

\bibitem[{\citenamefont{Srednicki}(1994)}]{Srednicki1994}
\bibinfo{author}{\bibfnamefont{M.}~\bibnamefont{Srednicki}},
  \bibinfo{journal}{Phys. Rev. E} \textbf{\bibinfo{volume}{50}},
  \bibinfo{pages}{888} (\bibinfo{year}{1994}).

\bibitem[{\citenamefont{Kinoshita et~al.}(2006)\citenamefont{Kinoshita, Wenger,
  and Weiss}}]{Kinoshita2006}
\bibinfo{author}{\bibfnamefont{T.}~\bibnamefont{Kinoshita}},
  \bibinfo{author}{\bibfnamefont{T.}~\bibnamefont{Wenger}}, \bibnamefont{and}
  \bibinfo{author}{\bibfnamefont{D.~S.} \bibnamefont{Weiss}},
  \bibinfo{journal}{Nature} \textbf{\bibinfo{volume}{440}},
  \bibinfo{pages}{900} (\bibinfo{year}{2006}).

\bibitem[{\citenamefont{Rigol}(2009)}]{Rigol2009}
\bibinfo{author}{\bibfnamefont{M.}~\bibnamefont{Rigol}},
  \bibinfo{journal}{Phys. Rev. Lett.} \textbf{\bibinfo{volume}{103}},
  \bibinfo{pages}{100403} (\bibinfo{year}{2009}).

\bibitem[{\citenamefont{Cassidy et~al.}(2011)\citenamefont{Cassidy, Clark, and
  Rigol}}]{Cassidy2011}
\bibinfo{author}{\bibfnamefont{A.~C.} \bibnamefont{Cassidy}},
  \bibinfo{author}{\bibfnamefont{C.~W.} \bibnamefont{Clark}}, \bibnamefont{and}
  \bibinfo{author}{\bibfnamefont{M.}~\bibnamefont{Rigol}},
  \bibinfo{journal}{Phys. Rev. Lett.} \textbf{\bibinfo{volume}{106}},
  \bibinfo{pages}{140405} (\bibinfo{year}{2011}).

\bibitem[{\citenamefont{Rigol and Fitzpatrick}(2011)}]{Rigol2011}
\bibinfo{author}{\bibfnamefont{M.}~\bibnamefont{Rigol}} \bibnamefont{and}
  \bibinfo{author}{\bibfnamefont{M.}~\bibnamefont{Fitzpatrick}},
  \bibinfo{journal}{Phys. Rev. A} \textbf{\bibinfo{volume}{84}},
  \bibinfo{pages}{033640} (\bibinfo{year}{2011}).

\bibitem[{\citenamefont{Bloch et~al.}(2008)\citenamefont{Bloch, Dalibard, and
  Zwerger}}]{Bloch2008}
\bibinfo{author}{\bibfnamefont{I.}~\bibnamefont{Bloch}},
  \bibinfo{author}{\bibfnamefont{J.}~\bibnamefont{Dalibard}}, \bibnamefont{and}
  \bibinfo{author}{\bibfnamefont{W.}~\bibnamefont{Zwerger}},
  \bibinfo{journal}{Rev. Mod. Phys.} \textbf{\bibinfo{volume}{80}},
  \bibinfo{pages}{885} (\bibinfo{year}{2008}).

\bibitem[{\citenamefont{Vidal}(2004)}]{Vidal2004}
\bibinfo{author}{\bibfnamefont{G.}~\bibnamefont{Vidal}},
  \bibinfo{journal}{Phys. Rev. Lett.} \textbf{\bibinfo{volume}{93}},
  \bibinfo{pages}{040502} (\bibinfo{year}{2004}).

\bibitem[{\citenamefont{Daley et~al.}(2004)\citenamefont{Daley, Kollath,
  Schollw\"ock, and Vidal}}]{Daley2004}
\bibinfo{author}{\bibfnamefont{A.~J.} \bibnamefont{Daley}},
  \bibinfo{author}{\bibfnamefont{C.}~\bibnamefont{Kollath}},
  \bibinfo{author}{\bibfnamefont{U.}~\bibnamefont{Schollw\"ock}},
  \bibnamefont{and} \bibinfo{author}{\bibfnamefont{G.}~\bibnamefont{Vidal}},
  \bibinfo{journal}{Journal of Statistical Mechanics: Theory and Experiment} p.
  \bibinfo{pages}{P04005} (\bibinfo{year}{2004}).

\bibitem[{\citenamefont{White and Feiguin}(2004)}]{White2004}
\bibinfo{author}{\bibfnamefont{S.~R.} \bibnamefont{White}} \bibnamefont{and}
  \bibinfo{author}{\bibfnamefont{A.~E.} \bibnamefont{Feiguin}},
  \bibinfo{journal}{Phys. Rev. Lett.} \textbf{\bibinfo{volume}{93}},
  \bibinfo{pages}{076401} (\bibinfo{year}{2004}).

\bibitem[{\citenamefont{Verstraete et~al.}(2008)\citenamefont{Verstraete, Murg,
  and Cirac}}]{Verstraete2008}
\bibinfo{author}{\bibfnamefont{F.}~\bibnamefont{Verstraete}},
  \bibinfo{author}{\bibfnamefont{V.}~\bibnamefont{Murg}}, \bibnamefont{and}
  \bibinfo{author}{\bibfnamefont{J.~I.} \bibnamefont{Cirac}},
  \bibinfo{journal}{Advances in Physics} \textbf{\bibinfo{volume}{57}},
  \bibinfo{pages}{143 } (\bibinfo{year}{2008}).

\bibitem[{\citenamefont{M{\o}lmer et~al.}(1993)\citenamefont{M{\o}lmer, Castin,
  and Dalibard}}]{Molmer1993}
\bibinfo{author}{\bibfnamefont{K.}~\bibnamefont{M{\o}lmer}},
  \bibinfo{author}{\bibfnamefont{Y.}~\bibnamefont{Castin}}, \bibnamefont{and}
  \bibinfo{author}{\bibfnamefont{J.}~\bibnamefont{Dalibard}},
  \bibinfo{journal}{J. Opt. Soc. Am. B} \textbf{\bibinfo{volume}{10}},
  \bibinfo{pages}{524} (\bibinfo{year}{1993}).

\bibitem[{\citenamefont{Gardiner and Zoller}(2005)}]{Gardiner2005}
\bibinfo{author}{\bibfnamefont{C.~W.} \bibnamefont{Gardiner}} \bibnamefont{and}
  \bibinfo{author}{\bibfnamefont{P.}~\bibnamefont{Zoller}},
  \emph{\bibinfo{title}{Quantum Noise}} (\bibinfo{publisher}{Springer, Berlin},
  \bibinfo{year}{2005}).

\bibitem[{\citenamefont{Carmichael}(1993)}]{Carmichael1993}
\bibinfo{author}{\bibfnamefont{H.~J.} \bibnamefont{Carmichael}},
  \emph{\bibinfo{title}{An Open Systems Approach to Quantum Optics}}
  (\bibinfo{publisher}{Springer, Berlin}, \bibinfo{year}{1993}).

\bibitem[{\citenamefont{McKay and DeMarco}(2011)}]{McKay2011}
\bibinfo{author}{\bibfnamefont{D.~C.} \bibnamefont{McKay}} \bibnamefont{and}
  \bibinfo{author}{\bibfnamefont{B.}~\bibnamefont{DeMarco}},
  \bibinfo{journal}{Reports on Progress in Physics}
  \textbf{\bibinfo{volume}{74}}, \bibinfo{pages}{054401}
  (\bibinfo{year}{2011}).

\bibitem[{\citenamefont{Fuchs et~al.}(2011)\citenamefont{Fuchs, Gull, Pollet,
  Burovski, Kozik, Pruschke, and Troyer}}]{Fuchs2011}
\bibinfo{author}{\bibfnamefont{S.}~\bibnamefont{Fuchs}},
  \bibinfo{author}{\bibfnamefont{E.}~\bibnamefont{Gull}},
  \bibinfo{author}{\bibfnamefont{L.}~\bibnamefont{Pollet}},
  \bibinfo{author}{\bibfnamefont{E.}~\bibnamefont{Burovski}},
  \bibinfo{author}{\bibfnamefont{E.}~\bibnamefont{Kozik}},
  \bibinfo{author}{\bibfnamefont{T.}~\bibnamefont{Pruschke}}, \bibnamefont{and}
  \bibinfo{author}{\bibfnamefont{M.}~\bibnamefont{Troyer}},
  \bibinfo{journal}{Phys. Rev. Lett.} \textbf{\bibinfo{volume}{106}},
  \bibinfo{pages}{030401} (\bibinfo{year}{2011}).

\bibitem[{\citenamefont{J\"ordens et~al.}(2010)\citenamefont{J\"ordens,
  Tarruell, Greif, Uehlinger, Strohmaier, Moritz, Esslinger, De~Leo, Kollath,
  Georges et~al.}}]{Jordens2010}
\bibinfo{author}{\bibfnamefont{R.}~\bibnamefont{J\"ordens}},
  \bibinfo{author}{\bibfnamefont{L.}~\bibnamefont{Tarruell}},
  \bibinfo{author}{\bibfnamefont{D.}~\bibnamefont{Greif}},
  \bibinfo{author}{\bibfnamefont{T.}~\bibnamefont{Uehlinger}},
  \bibinfo{author}{\bibfnamefont{N.}~\bibnamefont{Strohmaier}},
  \bibinfo{author}{\bibfnamefont{H.}~\bibnamefont{Moritz}},
  \bibinfo{author}{\bibfnamefont{T.}~\bibnamefont{Esslinger}},
  \bibinfo{author}{\bibfnamefont{L.}~\bibnamefont{De~Leo}},
  \bibinfo{author}{\bibfnamefont{C.}~\bibnamefont{Kollath}},
  \bibinfo{author}{\bibfnamefont{A.}~\bibnamefont{Georges}},
  \bibnamefont{et~al.}, \bibinfo{journal}{Phys. Rev. Lett.}
  \textbf{\bibinfo{volume}{104}}, \bibinfo{pages}{180401}
  (\bibinfo{year}{2010}).

\bibitem[{\citenamefont{Trotzky et~al.}(2010)\citenamefont{Trotzky, Pollet,
  Gerbier, Schnorrberger, Bloch, Prokof'ev, Svistunov, and
  Troyer}}]{Trotzky2010}
\bibinfo{author}{\bibfnamefont{S.}~\bibnamefont{Trotzky}},
  \bibinfo{author}{\bibfnamefont{L.}~\bibnamefont{Pollet}},
  \bibinfo{author}{\bibfnamefont{F.}~\bibnamefont{Gerbier}},
  \bibinfo{author}{\bibfnamefont{U.}~\bibnamefont{Schnorrberger}},
  \bibinfo{author}{\bibfnamefont{I.}~\bibnamefont{Bloch}},
  \bibinfo{author}{\bibfnamefont{N.~V.} \bibnamefont{Prokof'ev}},
  \bibinfo{author}{\bibfnamefont{B.}~\bibnamefont{Svistunov}},
  \bibnamefont{and} \bibinfo{author}{\bibfnamefont{M.}~\bibnamefont{Troyer}},
  \bibinfo{journal}{Nat Phys} \textbf{\bibinfo{volume}{6}},
  \bibinfo{pages}{998} (\bibinfo{year}{2010}).

\bibitem[{\citenamefont{Esslinger}(2010)}]{Esslinger2010}
\bibinfo{author}{\bibfnamefont{T.}~\bibnamefont{Esslinger}},
  \bibinfo{journal}{Annual Review of Condensed Matter Physics}
  \textbf{\bibinfo{volume}{1}}, \bibinfo{pages}{129} (\bibinfo{year}{2010}).

\bibitem[{\citenamefont{Bloch et~al.}(2012)\citenamefont{Bloch, Dalibard, and
  Nascimbene}}]{Bloch2012}
\bibinfo{author}{\bibfnamefont{I.}~\bibnamefont{Bloch}},
  \bibinfo{author}{\bibfnamefont{J.}~\bibnamefont{Dalibard}}, \bibnamefont{and}
  \bibinfo{author}{\bibfnamefont{S.}~\bibnamefont{Nascimbene}},
  \bibinfo{journal}{Nat Phys} \textbf{\bibinfo{volume}{8}},
  \bibinfo{pages}{267} (\bibinfo{year}{2012}).

\bibitem[{\citenamefont{Cirac and Zoller}(2012)}]{Cirac2012}
\bibinfo{author}{\bibfnamefont{J.~I.} \bibnamefont{Cirac}} \bibnamefont{and}
  \bibinfo{author}{\bibfnamefont{P.}~\bibnamefont{Zoller}},
  \bibinfo{journal}{Nat Phys} \textbf{\bibinfo{volume}{8}},
  \bibinfo{pages}{264} (\bibinfo{year}{2012}).

\bibitem[{\citenamefont{Strohmaier et~al.}(2010)\citenamefont{Strohmaier,
  Greif, J\"ordens, Tarruell, Moritz, Esslinger, Sensarma, Pekker, Altman, and
  Demler}}]{Strohmaier2010}
\bibinfo{author}{\bibfnamefont{N.}~\bibnamefont{Strohmaier}},
  \bibinfo{author}{\bibfnamefont{D.}~\bibnamefont{Greif}},
  \bibinfo{author}{\bibfnamefont{R.}~\bibnamefont{J\"ordens}},
  \bibinfo{author}{\bibfnamefont{L.}~\bibnamefont{Tarruell}},
  \bibinfo{author}{\bibfnamefont{H.}~\bibnamefont{Moritz}},
  \bibinfo{author}{\bibfnamefont{T.}~\bibnamefont{Esslinger}},
  \bibinfo{author}{\bibfnamefont{R.}~\bibnamefont{Sensarma}},
  \bibinfo{author}{\bibfnamefont{D.}~\bibnamefont{Pekker}},
  \bibinfo{author}{\bibfnamefont{E.}~\bibnamefont{Altman}}, \bibnamefont{and}
  \bibinfo{author}{\bibfnamefont{E.}~\bibnamefont{Demler}},
  \bibinfo{journal}{Phys. Rev. Lett.} \textbf{\bibinfo{volume}{104}},
  \bibinfo{pages}{080401} (\bibinfo{year}{2010}).

\bibitem[{\citenamefont{Kolovsky and Buchleitner}(2004)}]{0295-5075-68-5-632}
\bibinfo{author}{\bibfnamefont{A.~R.} \bibnamefont{Kolovsky}} \bibnamefont{and}
  \bibinfo{author}{\bibfnamefont{A.}~\bibnamefont{Buchleitner}},
  \bibinfo{journal}{EPL (Europhysics Letters)} \textbf{\bibinfo{volume}{68}},
  \bibinfo{pages}{632} (\bibinfo{year}{2004}).

\bibitem[{\citenamefont{Kolovsky}(2007)}]{PhysRevLett.99.020401}
\bibinfo{author}{\bibfnamefont{A.~R.} \bibnamefont{Kolovsky}},
  \bibinfo{journal}{Phys. Rev. Lett.} \textbf{\bibinfo{volume}{99}},
  \bibinfo{pages}{020401} (\bibinfo{year}{2007}).

\bibitem[{\citenamefont{Rigol et~al.}(2007)\citenamefont{Rigol, Dunjko,
  Yurovsky, and Olshanii}}]{Rigol2007}
\bibinfo{author}{\bibfnamefont{M.}~\bibnamefont{Rigol}},
  \bibinfo{author}{\bibfnamefont{V.}~\bibnamefont{Dunjko}},
  \bibinfo{author}{\bibfnamefont{V.}~\bibnamefont{Yurovsky}}, \bibnamefont{and}
  \bibinfo{author}{\bibfnamefont{M.}~\bibnamefont{Olshanii}},
  \bibinfo{journal}{Phys. Rev. Lett.} \textbf{\bibinfo{volume}{98}},
  \bibinfo{pages}{050405} (\bibinfo{year}{2007}).

\bibitem[{\citenamefont{Biroli et~al.}(2010)\citenamefont{Biroli, Kollath, and
  L\"auchli}}]{Biroli2010}
\bibinfo{author}{\bibfnamefont{G.}~\bibnamefont{Biroli}},
  \bibinfo{author}{\bibfnamefont{C.}~\bibnamefont{Kollath}}, \bibnamefont{and}
  \bibinfo{author}{\bibfnamefont{A.~M.} \bibnamefont{L\"auchli}},
  \bibinfo{journal}{Phys. Rev. Lett.} \textbf{\bibinfo{volume}{105}},
  \bibinfo{pages}{250401} (\bibinfo{year}{2010}).

\bibitem[{\citenamefont{Natu et~al.}(2011)\citenamefont{Natu, Hazzard, and
  Mueller}}]{Natu2011}
\bibinfo{author}{\bibfnamefont{S.~S.} \bibnamefont{Natu}},
  \bibinfo{author}{\bibfnamefont{K.~R.~A.} \bibnamefont{Hazzard}},
  \bibnamefont{and} \bibinfo{author}{\bibfnamefont{E.~J.}
  \bibnamefont{Mueller}}, \bibinfo{journal}{Phys. Rev. Lett.}
  \textbf{\bibinfo{volume}{106}}, \bibinfo{pages}{125301}
  (\bibinfo{year}{2011}).

\bibitem[{\citenamefont{Hung et~al.}(2010)\citenamefont{Hung, Zhang, Gemelke,
  and Chin}}]{Hung2010}
\bibinfo{author}{\bibfnamefont{C.-L.} \bibnamefont{Hung}},
  \bibinfo{author}{\bibfnamefont{X.}~\bibnamefont{Zhang}},
  \bibinfo{author}{\bibfnamefont{N.}~\bibnamefont{Gemelke}}, \bibnamefont{and}
  \bibinfo{author}{\bibfnamefont{C.}~\bibnamefont{Chin}},
  \bibinfo{journal}{Phys. Rev. Lett.} \textbf{\bibinfo{volume}{104}},
  \bibinfo{pages}{160403} (\bibinfo{year}{2010}).

\bibitem[{\citenamefont{Bakr et~al.}(2010)\citenamefont{Bakr, Peng, Tai, Ma,
  Simon, Gillen, F\"olling, Pollet, and Greiner}}]{Bakr2010}
\bibinfo{author}{\bibfnamefont{W.~S.} \bibnamefont{Bakr}},
  \bibinfo{author}{\bibfnamefont{A.}~\bibnamefont{Peng}},
  \bibinfo{author}{\bibfnamefont{M.~E.} \bibnamefont{Tai}},
  \bibinfo{author}{\bibfnamefont{R.}~\bibnamefont{Ma}},
  \bibinfo{author}{\bibfnamefont{J.}~\bibnamefont{Simon}},
  \bibinfo{author}{\bibfnamefont{J.~I.} \bibnamefont{Gillen}},
  \bibinfo{author}{\bibfnamefont{S.}~\bibnamefont{F\"olling}},
  \bibinfo{author}{\bibfnamefont{L.}~\bibnamefont{Pollet}}, \bibnamefont{and}
  \bibinfo{author}{\bibfnamefont{M.}~\bibnamefont{Greiner}},
  \bibinfo{journal}{Science} \textbf{\bibinfo{volume}{329}},
  \bibinfo{pages}{547} (\bibinfo{year}{2010}).

\bibitem[{\citenamefont{Prokof'ev et~al.}(1998)\citenamefont{Prokof'ev,
  Svistunov, and Tupitsyn}}]{Prokofev1998}
\bibinfo{author}{\bibfnamefont{N.}~\bibnamefont{Prokof'ev}},
  \bibinfo{author}{\bibfnamefont{B.}~\bibnamefont{Svistunov}},
  \bibnamefont{and} \bibinfo{author}{\bibfnamefont{I.}~\bibnamefont{Tupitsyn}},
  \bibinfo{journal}{Journal of Experimental and Theoretical Physics}
  \textbf{\bibinfo{volume}{87}}, \bibinfo{pages}{310} (\bibinfo{year}{1998}),
  ISSN \bibinfo{issn}{1063-7761}.

\bibitem[{\citenamefont{Pollet et~al.}(2007)\citenamefont{Pollet, Houcke, and
  Rombouts}}]{Pollet2007}
\bibinfo{author}{\bibfnamefont{L.}~\bibnamefont{Pollet}},
  \bibinfo{author}{\bibfnamefont{K.~V.} \bibnamefont{Houcke}},
  \bibnamefont{and} \bibinfo{author}{\bibfnamefont{S.~M.}
  \bibnamefont{Rombouts}}, \bibinfo{journal}{Journal of Computational Physics}
  \textbf{\bibinfo{volume}{225}}, \bibinfo{pages}{2249 }
  (\bibinfo{year}{2007}), ISSN \bibinfo{issn}{0021-9991}.

\bibitem[{\citenamefont{Pollet}(2012)}]{Pollet2012}
\bibinfo{author}{\bibfnamefont{L.}~\bibnamefont{Pollet}},
  \bibinfo{journal}{Reports on Progress in Physics}
  \textbf{\bibinfo{volume}{75}}, \bibinfo{pages}{094501}
  (\bibinfo{year}{2012}).

\bibitem[{\citenamefont{Toth and Blakie}(2011)}]{Toth2011}
\bibinfo{author}{\bibfnamefont{E.}~\bibnamefont{Toth}} \bibnamefont{and}
  \bibinfo{author}{\bibfnamefont{P.~B.} \bibnamefont{Blakie}},
  \bibinfo{journal}{Physical Review A} \textbf{\bibinfo{volume}{83}},
  \bibinfo{pages}{021601(R)} (\bibinfo{year}{2011}).

\bibitem[{\citenamefont{Daley et~al.}(2009)\citenamefont{Daley, Taylor, Diehl,
  Baranov, and Zoller}}]{Daley2009}
\bibinfo{author}{\bibfnamefont{A.~J.} \bibnamefont{Daley}},
  \bibinfo{author}{\bibfnamefont{J.~M.} \bibnamefont{Taylor}},
  \bibinfo{author}{\bibfnamefont{S.}~\bibnamefont{Diehl}},
  \bibinfo{author}{\bibfnamefont{M.}~\bibnamefont{Baranov}}, \bibnamefont{and}
  \bibinfo{author}{\bibfnamefont{P.}~\bibnamefont{Zoller}},
  \bibinfo{journal}{Phys. Rev. Lett.} \textbf{\bibinfo{volume}{102}},
  \bibinfo{pages}{040402} (\bibinfo{year}{2009}).

\bibitem[{\citenamefont{B\"uchler}(2010)}]{Buchler2010}
\bibinfo{author}{\bibfnamefont{H.~P.} \bibnamefont{B\"uchler}},
  \bibinfo{journal}{Phys. Rev. Lett.} \textbf{\bibinfo{volume}{104}},
  \bibinfo{pages}{090402} (\bibinfo{year}{2010}).

\bibitem[{\citenamefont{Mark et~al.}(2011)\citenamefont{Mark, Haller, Lauber,
  Danzl, Daley, and N\"agerl}}]{Mark2011}
\bibinfo{author}{\bibfnamefont{M.~J.} \bibnamefont{Mark}},
  \bibinfo{author}{\bibfnamefont{E.}~\bibnamefont{Haller}},
  \bibinfo{author}{\bibfnamefont{K.}~\bibnamefont{Lauber}},
  \bibinfo{author}{\bibfnamefont{J.~G.} \bibnamefont{Danzl}},
  \bibinfo{author}{\bibfnamefont{A.~J.} \bibnamefont{Daley}}, \bibnamefont{and}
  \bibinfo{author}{\bibfnamefont{H.-C.} \bibnamefont{N\"agerl}},
  \bibinfo{journal}{Phys. Rev. Lett.} \textbf{\bibinfo{volume}{107}},
  \bibinfo{pages}{175301} (\bibinfo{year}{2011}).

\bibitem[{\citenamefont{Mark et~al.}(2012)\citenamefont{Mark, Haller, Lauber,
  Danzl, Janisch, B\"uchler, Daley, and N\"agerl}}]{Mark2012}
\bibinfo{author}{\bibfnamefont{M.~J.} \bibnamefont{Mark}},
  \bibinfo{author}{\bibfnamefont{E.}~\bibnamefont{Haller}},
  \bibinfo{author}{\bibfnamefont{K.}~\bibnamefont{Lauber}},
  \bibinfo{author}{\bibfnamefont{J.~G.} \bibnamefont{Danzl}},
  \bibinfo{author}{\bibfnamefont{A.}~\bibnamefont{Janisch}},
  \bibinfo{author}{\bibfnamefont{H.~P.} \bibnamefont{B\"uchler}},
  \bibinfo{author}{\bibfnamefont{A.~J.} \bibnamefont{Daley}}, \bibnamefont{and}
  \bibinfo{author}{\bibfnamefont{H.-C.} \bibnamefont{N\"agerl}},
  \bibinfo{journal}{Phys. Rev. Lett.} \textbf{\bibinfo{volume}{108}},
  \bibinfo{pages}{215302} (\bibinfo{year}{2012}).

\bibitem[{\citenamefont{Carmichael}(1991)}]{carmichael_open_1991}
\bibinfo{author}{\bibfnamefont{H.}~\bibnamefont{Carmichael}},
  \emph{\bibinfo{title}{An Open Systems Approach to Quantum Optics, Lectures
  Presented at the Universit\'e Libre de Bruxelles}}, Lecture Notes in Physics
  monographs (\bibinfo{publisher}{Springer}, \bibinfo{year}{1991}).

\bibitem[{\citenamefont{Molmer et~al.}(1993)\citenamefont{Molmer, Castin, and
  Dalibard}}]{molmer_monte_1993}
\bibinfo{author}{\bibfnamefont{K.}~\bibnamefont{Molmer}},
  \bibinfo{author}{\bibfnamefont{Y.}~\bibnamefont{Castin}}, \bibnamefont{and}
  \bibinfo{author}{\bibfnamefont{J.}~\bibnamefont{Dalibard}},
  \bibinfo{journal}{Journal of the Optical Society of America B}
  \textbf{\bibinfo{volume}{10}}, \bibinfo{pages}{524} (\bibinfo{year}{1993}).

\bibitem[{\citenamefont{Dum et~al.}(1992)\citenamefont{Dum, Parkins, Zoller,
  and Gardiner}}]{dum_monte_1992}
\bibinfo{author}{\bibfnamefont{R.}~\bibnamefont{Dum}},
  \bibinfo{author}{\bibfnamefont{A.~S.} \bibnamefont{Parkins}},
  \bibinfo{author}{\bibfnamefont{P.}~\bibnamefont{Zoller}}, \bibnamefont{and}
  \bibinfo{author}{\bibfnamefont{C.~W.} \bibnamefont{Gardiner}},
  \bibinfo{journal}{Physical Review A} \textbf{\bibinfo{volume}{46}},
  \bibinfo{pages}{4382} (\bibinfo{year}{1992}).

\end{thebibliography}

\end{document}